\documentclass{article}

\usepackage[final]{nips_2018}

\usepackage[utf8]{inputenc} % allow utf-8 input
\usepackage[T1]{fontenc}    % use 8-bit T1 fonts
\usepackage{hyperref}       % hyperlinks
\usepackage{url}            % simple URL typesetting
\usepackage{booktabs}       % professional-quality tables
\usepackage{amsfonts}       % blackboard math symbols
\usepackage{nicefrac}       % compact symbols for 1/2, etc.
\usepackage{microtype}      % microtypography
\usepackage{graphicx}
\usepackage[cmex10]{amsmath}
\title{An adaptive treatment recommendation and outcome prediction model for metastatic melanoma}

\author{
  Xue Teng\\
  Pulse Infoframe Inc., London, ON  \\
Schulich School of Medicine \& Dentistry\\ University of Western Ontario, London, ON \\
  \texttt{sue.teng@ieee.org} \\
\And
  Fuad Gwadry\\
  Pulse Infoframe Inc.\\
  London, ON \\
  \texttt{fgwadry@pulseinfoframe.com} \\
\And
  Haley McConkey\\
  Global Melanoma Research Network\\
  London, ON \\
  \texttt{hmcconkey@wwmrn.org} \\
\And
  Scott Ernst\\
  Schulich School of Medicine \& Dentistry\\ University of Western Ontario, London, ON \\
  \texttt{Scott.Ernst@lhsc.on.ca} \\
\And
  Femida Gwadry-Sridhar\\
  Department of Computer Science\\
  University of Western Ontario, London, ON \\
  \texttt{fgwadrys@uwo.ca} \\
}

\begin{document}

\maketitle

\begin{abstract}
Melanoma is a type of skin cancer developed from melanocytes. It is one of the most lethal types of cancer, accounting for approximately 75\% of skin cancer deaths. Late stage melanoma is very difficult to treat, since the cancer cells are deranged, may be genetically linked and can be unresponsive to therapy. Therefore, determining how to effectively make use of different treatment regimens is of vital importance to survival. In this analysis, we propose an adaptive treatment recommendation system based on a hybrid cluster-classification (CC) structure. Our proposed system consists of two parts,1) distribution based clustering and 2) classification. Our recommendation system can help to identify high-risk melanoma patients and suggest the best approach to treatment, which enables clinicians and patients to make decisions on the basis of real-world data. Our data came from the Canadian Melanoma Research Network (CMRN) database, a pan-Canadian multi-year observational database, which is part of Global Melanoma Registry Network (GMRN). Training/testing sets are generated based on data from different sources, leading to cross cohort analysis tasks. Experimental results show that our proposed system achieves very promising results with an overall accuracy of up to 80\%.
\end{abstract}

\section{Introduction}

Melanoma is the most dangerous type of skin cancer, accounting for only 2\% of all skin cancer types, yet approximately 75\% of skin cancer deaths in Canada \cite{ca_stats}. Incidence rates of melanoma have increased in both men and women over the past several decades \cite{cancer2015}. Early stages of melanoma (no spread) can mostly be cured, while late stage melanoma (with spread) are very difficult to treat. Once metastasized, the cancer cells impose tremendous burden to the human body and often resist treatment \cite{Wu2012}. Melanoma is typically caused by DNA damage resulting from exposure to ultraviolet radiation. Cancerous cells then  develop from the melanin-producing cells, known as melanocytes that obtain genetic alterations that promote unregulated proliferation. The proliferation of melanocytes can either cause the formation of melanocytic nevus or melanoma. Early stage melanoma (without spread) can be cured by surgical  excision, while late stage melanoma requires a wide variety of treatments including radiation, surgery, immune-oncology agents, targeted therapy and chemotherapy. 
 
In clinical practice, treatment decisions are usually made based on physician experience which they supplement using information available from clinical trials and objective diagnostic measures. In the past, machine learning algorithms have been used for the diagnosis of melanoma based on dermoscopy images; however, less effort has been directed on the analysis of treatment outcomes or treatment planning using advanced machine learning algorithms. Studies continue to use traditional statistical methods, such as logistic regression andsurvival analysis,  which are limited in their ability to provide deeper insight into why certain outcomes occur in certain patients. The recent advent of multiple treatment options for patients has led to variable outcomes that may be a result of various factors such as age, gender, melanoma location, size, thickness, etc. As medicine moves towards a more personalized approach , it has become increasingly clear that there is a need to explore different machine learning methodologies to determine whether they can help supplement clinical decisions and incorporate patient preferences to improve patient outcomes. 

We propose a general framework for the analysis of melanoma treatment plans. In particular, we present a cluster-classification (CC) structure to provide optimized treatment recommendations and outcomes prediction. We combine distribution based clustering algorithm with support vector machine (SVM) to provide the optimal treatment sequence in a Bayesian framework.

\section{Background}
The fast development of machine learning provides valuable tools for analysis of heterogeneous medical data \cite{Dai2017}.  Machine learning has been widely used in medical research. Identification of mutations in unexplored genome regions is among the most popular topics \cite{Zhou2015}. Another important area is medical image processing, which provides valuable insights about the hidden information \cite{Dai2016}. Farina et al. explored the potential and limits of multispectral imaging approach in the diagnosis of cutaneous melanoma using telespectrophotometric technique \cite{Farina2000}. Their results suggested the possibility  that a  computer-aided device could discriminate malignant from benign lesions, though the system is relatively simple.  

Recently, treatment planning using predictive analytics has begun to draw more and more attention. Moore et al. constructed a SQL database for initial dosage prediction \cite{Moore2014}. Li et al. proposed an automatic system for Adaptive radiation therapy (ART) to replace the manual trial-and-error process \cite{Li2013}. However, for melanoma, automatic treatment planning is still a relatively new concept. To the best of our knowledge, this is the first work which systematically discusses the application of machine learning techniques into treatment plan optimization and outcome prediction. Transfering knowledge between different study cohort has also become a very effective approach \cite{DaiKBS}.

\section{Methods}
\begin{figure}
  \centering
  \includegraphics[clip=true,viewport= 0 230 950 550, width=0.95\textwidth]{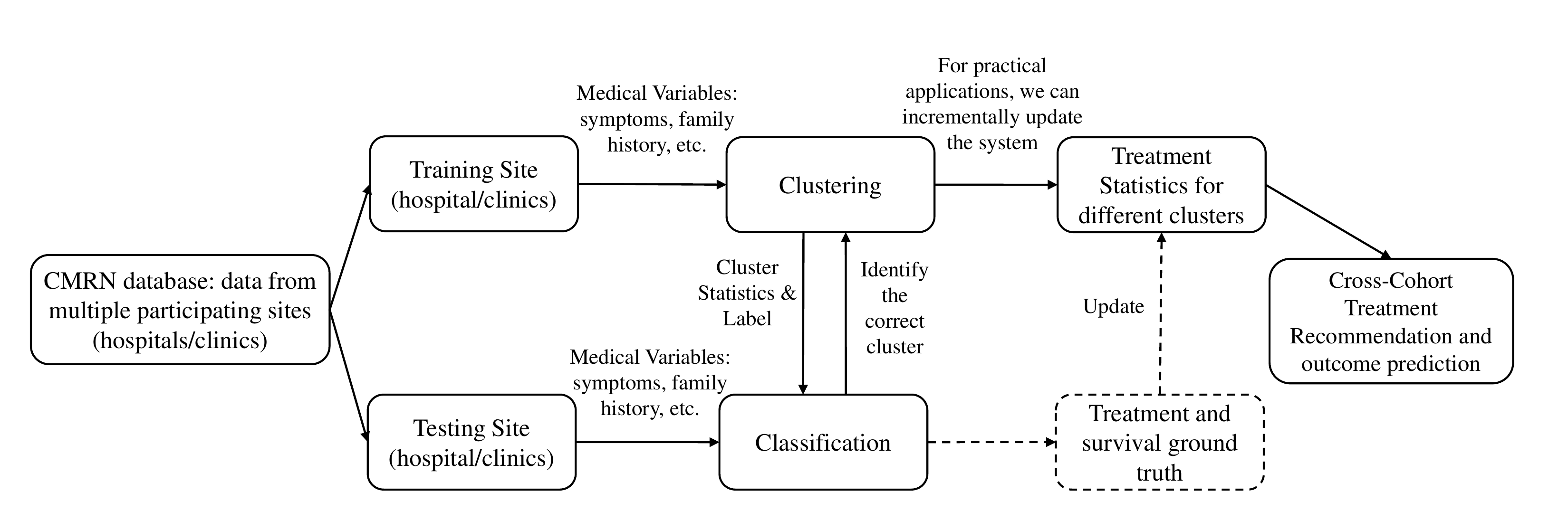}
  \caption{Schematic overview of the proposed methodology.}
  \label{fig:system}
\end{figure}

The goal of the proposed research is to build a generalized treatment recommendation and outcome prediction system. Figure \ref{fig:system} presents a diagram outliningthe proposed system. The general idea is to first cluster the patients into three broad categories corresponding to different stages of melanoma. Then, statistical analysis is utilized to extract the characteristics of different clusters, which is used to provide treatment recommendations, e.g. treatment regimens and sequencing. In addition, our proposed system can adapt to specific cohort in practical application. In Figure \ref{fig:system}, the testing data (corresponding to new patients in practical applications) can contribute to the survival statistics. Thus, we can update the system recurrently, which enables our system to adapt to a specific study cohort.

In our proposed framework, we first cluster the patients based on the clinical and demographic variables, such as melanoma size, thickness, location, and family history. There are up to 10 months mean survival time difference between different clusters. In our current implementation, a Gaussian Mixture Model (GMM) is utilized for clustering. The support vector machine (SVM) algorithm is adopted for classification. It has to be noted that our objective is to present a general framework for adaptive melanoma treatment recommendation. SVM can be replaced by other algorithms if desired. 

With the new patient medical variable vector, $\bf V$, and the corresponding cluster, $C$, we calculate the distance of $\bf V$ to all the records, $\bf M$, in C using the Euclidean distance (L2 norm). 
Thus, we obtain a ranking list of similar patients and the corresponding treatment plans as well as the treatment outcomes. The recommendation is made by a weighted voting among the top N most similar patients by
$
R_{regimen} = \frac{\sum (\gamma_k I_k)}{\sum (\gamma_k}
$, 
where $
I_k = \left\{\begin{matrix}
1, & {\bf M}_k \in top\ N\\ 
0, & {\bf M}_k \notin top\ N
\end{matrix}\right.
$. 
Thus the results are 
\begin{equation}\label{eq:lpp2}
R_{rec}=\max(R_1,R_2, \cdots)
\end{equation}
where $R_i$ corresponds to the scores for different regimens, i.e. Ipilimumab, Chemotherapy, Immune Inhibitors, and Vemurafenib.

\section{Results and Analysis}

\subsection{Experiment Database}

The clinical and treatment data from the Global  Melanoma Registry Network (GMRN) are used for verification tests \cite{gmrn}. The Canadian Melanoma Research Network (CMRN) is part of the GMRN, which consists of multiple participating sites, i.e. hospitals or clinics. CMRN attempts to capture all aspects of melanoma treatment, which requires the participating sites to enter data that is both evidence-based and relevant to the practice of melanoma specialists. This clinical dataset is then agreed to be entered into a common repositoryand all participating sites contribute by entering data. This leads to a wide variety of medical variables. Another important feature of CMRN is the large study cohort. In order to ascertain accurate results, each site is required to enter a minimum of 25 patient records per month. In our current implementation, we utilize the data from 6 participating sites. There are totally 1486 available metastatic regimen records and 584 metastatic patients with valid metastatic treatment record, 388 patients received ipilimumab therapy, 373 patients have multiple treatments. There are 126 patients with BRAF mutation positive, 43 patients with BRAF mutation negative and 415 patients without BRAF mutation information. 107 out of 126 patients with BRAF mutation positive received vemurafenib.

\subsection{Clustering}
Late stage melanoma (metastatic) has an extremely high death rate. For example, the respective 1, 2, and 5 year survival rates for Stage IV visceral melanoma are 33\%, 18\%, and 9\% \cite{melanoma1}. A typical treatment plan usually involves the induction of multiple regimens. Sequencing of regimens becomes one of the most important problems. There is a significant difference between the treatment procedures of various stages of melanoma. Therefore, for best recommendation results, it is better to narrow down the analysis to similar patients. In our current implementation, we utilize distribution based clustering algorithm for preprocessing.
We cluster the patients into 3 broad categories, corresponding to different symptom groups of melanoma. It has to be noted that there are clear survival differences between different clusters.

The mean survival times for different clusters are 11.75, 22.42, and 14.49 months. Therefore, the second cluster, denoted as red stars, shows the best survival expectation. A close investigation reveals that patients in cluster 2 generally are in earlier stages of melanoma, e.g. no lymphatic invasion (72.19\%), lower Clark’s level (89.47\% in stage I), no perineural invasion (72.54\%), etc. It has to be noted that clustering algorithm works on the high level similarity between different patients rather than individual similarity. In other words, the patient may suffer from lymphatic invasion but relatively good symptoms in terms of other criteria. This can be validated by the categorized survival statistics given in Table \ref{tab:surv}.

\begin{table}
  \caption{Survival statistics for different pathology (months).}
  \label{tab:surv}
  \centering
  \begin{tabular}{llllll}
\hline
 & Ulceration & Lymphatic & Perinueral & Microscopic & Lymphocytic \\
 & & Invasion & Invasion & Satelliotosis & Response \\
\hline
Cluster 1 & 21.6 & 17.89 & 9.04 & 6.66 & 11.6\\
Cluster 2 & 23.07 & 22.97 & 26.19 & 26.75 & 24.38\\
Cluster 3 & 18.57 & 18.89 & 16.81 & 11.94 & 15.61\\
\hline
\end{tabular}
\end{table}

\subsection{Treatment Recommendation}
The previous section gives a general idea of how different regimens contribute to the treatment results. The results are twofold. On the one hand, a good treatment outcome, e.g. high survival rate, indicates superior performance of the regimen. On the other hand, certain regimens are mainly used for late stage melanoma, which makes it unlikely to obtain high survival observations. Therefore, clustering algorithm is used to preprocess the data, so that only patients with similar symptoms are used for prediction. Table \ref{tab:recommendation} gives a typical recommendation output. It includes high risk symptoms, suggested treatment, suggested treatment sequence, and expected survival statistics. All those statistics are generated based on the patient cluster (e.g. \#2) and top N (e.g. N=10) similarity list. Upon special request, we can also generate other statistical analysis results, e.g. the Kaplan-Meier survival curves, Cox survival model, etc. 

The clinician’s treatment decision is used as the ground truth. Thus, we can calculate the prediction accuracy. Table \ref{tab:prediction} gives the treatment prediction/recommendation results. There are in total 6 participating sites in the CMRN database. We choose 5 sites as the training data and the rest as the testing data. The clinical decisions in the testing data are used as the ground truth. Table \ref{tab:prediction} gives the treatment recommendation (prediction) results. Our proposed system yields very promising results for the 1st treatment regimen prediction. However, when it comes to other line treatment regimens, there is large drop in the results. This is mainly due to the fact that a large amount of patients do not have a different treatment regimen in the treatment sequence. It has to be noted that Cluster 2 shows slightly better accuracy in 2nd line prediction. This may be caused by the longer survival time, which brings the patients and clinicians more options for treatment. The slight rise in prediction accuracy in 3rd line \& later regimen prediction accuracy is due to the increased use of Ipilimumab in late stage melanoma.
Another thing that has to be mentioned is that with additional information, e.g. manual labeling of melanoma stage by clinical experts, we can replace the clustering step by supervised learning algorithms, such as deep learning neural networks. Then, the ‘group’ can be more accurate, which in return improves the cluster statistics for treatment recommendation.

\begin{table}
  \caption{Typical recommendation output.}
  \label{tab:recommendation}
  \centering
  \begin{tabular}{ll}
\hline
Patient cluster: \#2 (83\%) & \\
\hline
High risk symptoms & Ulceration, Lymphatic Invasion Unknown \\
Suggested treatment & Chemotherapy, 1 Year Survival-62.62\%, 2 Year Survival-46.73\% \\
Suggested sequencing & Chemotherapy, Chemotherapy re-induction, Ipilimumab \\
Expected survival time & 36 months (std=2.1 months) \\

\hline
\end{tabular}
\end{table}

\begin{table}
  \caption{Treatment prediction results, (prediction options: Ipilimumab, Vemurafenib, Immune Inhibitors, Chemotherapy).}
  \label{tab:prediction}
  \centering
  \begin{tabular}{llll}
\hline
 & 1st line regimen & 2nd line regimen & 3rd line and later \\
\hline
Cluster 1 &75.37\% & 47.21\% & 50.12\%\\
Cluster 2 &78.31\% & 55.13\% & 55.89\%\\
Cluster 3 &79.01\% & 38.95\% & 40.27\%\\
\hline
\end{tabular}
\end{table}

\section{Conclusion and Discussion}
In this paper we propose a general framework for treatment recommendation, which possesses a hybrid cluster-classification (CC) structure. The treatment recommendation and outcome statistics are generated by machine-learning experience gained from similar patients. 
Our system can help = identify high-risk melanoma patients and provide recommendations, which allows clinicians and patients to make decisions regarding the best opportunities for patient responses to treatment . In our current implementation the system does not require any  prior clinical knowledge about melanoma. The system works in an unsupervised manner. It has to be noted that system performance can be improved by the incorporation of clinical knowledge. For example, manual labeling of disease stage or high risk patients can help to improve the system performance. The importance of disease specific data that generally does not exist in electronic medical records is key to helping specialist physicians and researchers advance treatments and improve patient outcomes. Machine learning provides a unique opportunity to better utilize available  data as a more precise decision aid.

\bibliographystyle{plain}
%\bibliography{ref_uwo}

\end{document}